\pdfoutput=1
\documentclass[11pt]{article}

\usepackage{csquotes}

\usepackage[round,longnamesfirst]{natbib}
\usepackage{amsmath, amsthm, amssymb, amsbsy,enumitem}
\setitemize{noitemsep}
\usepackage{array}
\usepackage{graphicx}
\usepackage{multirow}
\usepackage{cancel}
\usepackage[noend]{algorithmic}
\usepackage{algorithm}
\usepackage{setspace}
\usepackage{url}
\usepackage{verbatim}

\usepackage[bookmarksnumbered=true,backref=page,hyperfigures=true]{hyperref}

\usepackage{ifthen}
\usepackage{color}

\def\y{{\boldsymbol{y}}}
\def\Y{{\boldsymbol{Y}}}
\def\covariate{{\boldsymbol{x}}}

\def\meanstats{{\boldsymbol{\mu}}}

\def\natmap{{\boldsymbol{\eta}}}
\def\curvpar{{\boldsymbol{\theta}}}

\def\genstat{{\boldsymbol{g}}}
\def\gradient{{\boldsymbol{G}}}
\def\target{{\boldsymbol{t}}}
\def\V{{\boldsymbol{V}}}
\def\nactors{n}
\def\actors{N}

\def\form{^+}
\def\diss{^-}

\def\changeij{{\boldsymbol{\Delta}\sij}}
\def\setsub{\backslash}

\DeclareMathOperator{\E}{E}
\DeclareMathOperator{\age}{age}
\DeclareMathOperator{\Var}{Var}
\DeclareMathOperator{\logit}{logit}

\DeclareMathOperator{\Geometric}{Geometric}

\def\dysY{\mathbb{Y}}
\def\netsY{\mathcal{Y}}

\DeclareMathOperator{\Uniform}{Uniform}
\DeclareMathOperator{\Prob}{Pr}
\def\Peg{\Prob_{\natmap,\genstat}}
\def\PegF{\Prob_{\natmap\form,\genstat\form}}
\def\PegD{\Prob_{\natmap\diss,\genstat\diss}}
\def\Eeg{\E_{\natmap,\genstat}}
\def\Veg{\Var_{\natmap,\genstat}}
\DeclareMathOperator{\Odds}{Odds}
\def\Oeg{\Odds_{\natmap,\genstat}}
\def\ceg{c_{\natmap,\genstat}}
\def\cegF{c_{\natmap\form,\genstat\form}}
\def\cegD{c_{\natmap\diss,\genstat\diss}}
\def\netsYF{\mathcal{Y}\form}
\def\netsYD{\mathcal{Y}\diss}
\def\YF{\Y\form}
\def\YD{\Y\diss}
\def\yF{\y\form}
\def\yD{\y\diss}
\def\genstatF{\genstat\form}
\def\genstatD{\genstat\diss}
\def\curvparF{\curvpar\form}
\def\curvparD{\curvpar\diss}
\def\natmapF{\natmap\form}
\def\natmapD{\natmap\diss}
\def\argmax{\arg\max}

\DeclareMathOperator{\ilogit}{logit^{-1}}

\def\NN{\mathbb{N}}
\def\pij{(i,j)}
\def\pipjp{(i,j)}

\def\ijdysY{\pij\in\dysY}
\def\iactors{i\in\actors}
\def\ynetsY{\y\in\netsY}
\def\ypnetsY{\y\in\netsY}
\def\sij{_{i,j}}

\def\Yij{\Y\sij}
\def\yij{\y\sij}

\def\Yyij{\Yij=\yij}
\def\Yy{\Y=\y}
\def\half{\frac{1}{2}}

\def\natpar{\natmap(\curvpar)}
\def\natparF{\natmap\form(\curvpar\form)}
\def\natparD{\natmap\diss(\curvpar\diss)}
\def\curvparF{\curvpar\form}
\def\curvparD{\curvpar\diss}

\def\ErRe{Erd\H{o}s-R\'{e}nyi}

\newcommand{\myexp}[1]{\exp\left(#1\right)}

\newcommand{\I}[1]{1_{#1}}
\newcommand{\pkg}[1]{\textbf{#1}}
\newcommand{\proglang}[1]{\textsf{#1}}

\newcommand{\yat}[1]{\y^{t#1}}
\newcommand{\Yat}[1]{\Y^{t#1}}
\newcommand{\Yyat}[1]{\Yat{#1}=\yat{#1}}
\newcommand{\Yya}[1]{\Y^{#1}=y^{#1}}
\newcommand{\yatij}[1]{\y^{t#1}\sij}
\newcommand{\Yatij}[1]{\Y^{t#1}\sij}
\newcommand{\Yyatij}[1]{\Yatij{#1}=\yatij{#1}}

\providecommand{\abs}[1]{\left\lvert#1\right\rvert}

\def\t{^{\top}}

\graphicspath{{graphics/}{common/}}

\title{Modeling of Dynamic Networks based on Egocentric Data with Durational Information}

\author{Pavel N. Krivitsky\\\texttt{p.krivitsky@unsw.edu.au}\\School of Mathematics and Statistics\\University of New South Wales\\Sydney, NSW, Australia}

\date{April 2012 (original)\\
\today{} (repost)}

\begin{document}
\maketitle
\begin{quotation}
  \small
  \noindent\begin{center}\textbf{History}\end{center}
  
  This preprint was originally published to Penn State University Department of Statistics web site as \emph{Technical Report 12--01} in April 2012. It was subsequently lost, along with others, in a web site migration. In order to return it to the public record, we are reposting it, unmodified except as noted here:
  \begin{itemize}
  \item Penn State Statistics technical report title page has been replaced.
  \item Baseline font size has been enlarged for readability.
  \item Author affiliation and contact information has been added.
  \item Some items in the bibliography have been reformatted or updated.
  \item These changes may affect pagination.
  \end{itemize}
  This version may be superseded by other versions in the future.
\end{quotation}

\clearpage\begin{abstract}
  Modeling of dynamic networks --- networks that evolve over time ---
  has manifold applications in many fields. In epidemiology in
  particular, there is a need for data-driven modeling of human sexual
  relationship networks for the purpose of modeling and simulation of
  the spread of sexually transmitted disease. Dynamic network data
  about such networks are extremely difficult to collect, however, and
  much more readily available are egocentrically sampled data of a
  network at a single time point, with some attendant information
  about the sexual history of respondents.

  \citet{KrHa14s} proposed a Separable Temporal ERGM (STERGM)
  framework, which facilitates separable modeling of the tie duration
  distributions and the structural dynamics of tie formation. In this
  work, we apply this modeling framework to this problem, by studying
  the long-run properties of STERGM processes, developing methods for
  fitting STERGMs to egocentrically sampled data, and extending the
  network size adjustment method of \citet{krivitsky2011ans} to
  dynamic models.
\end{abstract}

\section{Introduction}
Modeling of dynamic networks --- networks that evolve over time ---
has applications in many fields. In epidemiology in particular, there
is a need for data-driven modeling of human sexual relationship
networks for the purpose of modeling and simulation of the spread of
sexually transmitted disease. As \citet{morris1997cps} show, spread of
such disease is affected not just by the momentary number of
partnerships, but by their timing. To that end, the models used must
have realistic temporal structure as well as cross-sectional
structure.

Exponential-family random graph ($p^*$) models (ERGMs) for social
networks are a natural way to represent dependencies in
cross-sectional graphs and dependencies between graphs over time,
particularly in a discrete context, and \citet{robins2001rgm} first
described this approach. \citet{hanneke2010dtm} also define and
describe what they call a Temporal ERGM (TERGM), postulating an
exponential family for the transition probability from a network at
time $t$ to a network at time $t+1$.

\citet{holland1977dms}, \citet{frank1991sac}, and others describe
\emph{continuous-time} Markov models for evolution of social networks
\citep{doreian1997esn}, and the most popular parametrization is the
\emph{actor-oriented} model described by \citet{snijders2005mln},
which can be viewed in terms of actors making decisions to make and
withdraw ties to other actors.

Arguing that \enquote{social processes and factors that result in ties
  being formed are not the same as those that result in ties being
  dissolved}, \citet{KrHa14s} introduced a separable
formulation of discrete-time models for network evolution,
parametrized in terms of a process that controls formation of new ties
and a process that controls dissolution of extant ties, in which both
processes are (usually different) ERGMs --- a Separable Temporal ERGM
(STERGM). Thus, the model separates the factors that affect
\emph{incidence} of ties --- the rate at which new ties are formed ---
from their \emph{duration} --- how long they tend to last once they
do.

Most of the attention in modeling of dynamic networks has focused on
fitting the model to a network series
\citep{snijders2001ses,hanneke2010dtm,KrHa14s} or an
enumeration of instantaneous events among actors in the network
\citep{butts2008ref}. In the former case, the dyad census of the
network of interest is observed at multiple time points. In the latter
case, each event of interest and its exact time of occurrence is
observed. However, observing a social network of interest at
multiple time points is often difficult or even impossible.

In the case of sexual partnership networks, even observing a full
census of all ties among the actors of interest is rare. One example
is the cross-sectional study focusing on at-risk populations of
Colorado Springs, Colorado --- female sex workers and their sexual
partners, and injecting drug users and their partners --- by
\citet{woodhouse1994msn} and \citet{klovdahl1994sni}. With 595
individuals ultimately interviewed, a dyad census among those
individuals was observed, but a total of 5162 individuals were named
as contacts by the respondents (including the respondents themselves),
so the dyad census of the network is far from complete.
\citet{helleringer2007sns} came close to observing a dyad census of
sexual partnerships within a geographic region: they took a census of
residents of Likoma Island, Malawi, interviewing those aged 18--35
about their sexual partnerships, including \emph{names and other
  personally identifying information about their partners}, and then
matching the reported partners to their list of island residents. This
approach might not be practical in areas less isolated and with
greater population densities, and presents severe confidentiality
issues, limiting access to such data. Even these studies produced only
a network at a single time point, rendering the above-mentioned
methods unsuitable.

More common are egocentrically-sampled network data --- data
comprising information about respondents (\emph{egos}) and their
immediate partners (\emph{alters}) --- are much easier to collect and
may contain temporal information about the network ties, in the form
of each respondent's past history and (right-censored) duration of
ongoing ties. Examples include the National Health and Social Life
Survey (NHSLS) \citep{laumann1994sos} and Wave III of the National
Longitudinal Study of Adolescent Health (Add Health)
\citep{harris2003nls}. (Notably, the former dataset is publicly
available for download with no restrictions.) \citet{krivitsky2011ans}
described a technique for fitting cross-sectional ERGMs to
egocentrically sampled data, applying it to NHSLS.

In this work, we approach the problem of fitting dynamic models based
on limited, cross-sectional or egocentric, network data by modeling
the observed network cross-section as a long-run product of the
dynamic network process being modeled: its stationary
distribution. Focusing on the STERGMs of \citet{KrHa14s}, we
derive their long-run properties and propose a generalized method of
moments (GMME) estimation technique for fitting these networks to
available data.

The rest of this paper proceeds as follows. In Section~\ref{sec:sep},
we review the Separable Temporal ERGMs, and in
Section~\ref{sec:equil}, we study their long-run, stationary behavior
of TERGMs in general and STERGMs in particular. Based on that, in
Section~\ref{sec:est}, we develop a best-effort approach to fit
dynamic models to cross-sectional networks with durational information
using Generalized Method of Moments Estimation (GMME) applied to the
stationary distribution of the dynamic network model. In
Section~\ref{sec:est-ego}, we show that the Conditional MLE (CMLE)
methods of \citet{KrHa14s} and others cannot be fit to
egocentrically observed network data, even if these data contain
fairly detailed temporal information, necessitating EGMME in that case
as well. Finally, we show how the network size adjustment of
\citet{krivitsky2011ans} can be applied to construct
network-size-invariant dynamic models in
Section~\ref{sec:dyn-netsize}, and demonstrate our development on
sexual partnership data in Section~\ref{sec:example}.
 
\section{\label{sec:sep}Separable temporal ERGM}
We now review the model proposed by \citet{KrHa14s}. Using
their notation, let $\actors$ be the set of $\nactors=\abs{\actors}$
actors of interest, labeled $1,\dotsc,\nactors$, and let
$\dysY\subseteq\{\{i,j\}: \pij\in\actors\times\actors \land i\ne j\}$
be the set of dyads (potential partnerships) among the actors. (Thus,
the networks we model are undirected and have no self-loops, but,
unless otherwise noted, our results apply equally to directed networks
where $\dysY\subseteq\{(i,j): \pij\in\actors\times\actors \land i\ne
j\}$.) $\dysY$ may be a proper subset: for example, if only
heterosexual ties are being modeled. Then, the set of possible
networks $\netsY$ is the power set of dyads, $2^\dysY$. For a network
at time $t-1$, $\yat{-1}$, \citet{KrHa14s} define
$\netsYF(\yat{-1})= \{\y\in 2^\dysY:\y\supseteq\yat{-1}\}$ be the set
of networks that can be constructed by forming zero or more ties in
$\yat{-1}$ and $\netsYD(\yat{-1})= \{\y\in
2^\dysY:\y\subseteq\yat{-1}\}$ be the set of networks that can be
constructed by dissolving zero or more ties in $\yat{-1}$.

Given $\yat{-1}$, the network $\Yat{}$ at time $t$ is modeled as a
consequence of some ties being formed according to a conditional ERGM
\begin{equation*}\PegF(\YF = \yF|\Yyat{-1};\curvparF )=\frac{\myexp{\natparF\cdot \genstatF(\yF,\yat{-1})}}{\cegF(\curvparF ,\yat{-1})},\ \yF\in\netsYF(\yat{-1})\end{equation*}
specified by model parameters $\curvparF$, sufficient statistic
$\genstatF$, and, optionally, a canonical mapping $\natmapF$; and some
dissolved according to a conditional ERGM
\begin{equation*}\PegD(\YD = \yD|\Yyat{-1};\curvparD )=\frac{\myexp{\natparD\cdot \genstatD(\yD,\yat{-1})}}{\cegD(\curvparD ,\yat{-1})},\ \yD\in\netsYD(\yat{-1}),\end{equation*} 
specified by (usually different) $\curvparD$, $\genstatD$, and
$\natmapD$. Their normalizing constants $\cegF(\curvparF ,\yat{-1})$
and $\cegD(\curvparD ,\yat{-1})$ sum their respective model kernels
over $\netsYF(\yat{-1})$ and $\netsYD(\yat{-1})$, respectively.
$\Yat{}$ is then evaluated by applying the changes in $\YF$ and $\YD$ to $\yat{-1}$: $\Yat{}=\yat{-1}\cup (\yF\setsub\yat{-1})\setsub
(\yat{-1}\setsub\yD)=\yF\setsub(\yat{-1}\setsub\yD)=\yD\cup(\yF\setsub\yat{-1})$.
The transition probability from $\yat{-1}$ to $\Yat{}$ is itself an ERGM, making STERGM a submodel of the Temporal ERGM (TERGM) of \citet{hanneke2010dtm}:
\begin{equation}
  \Peg(\Yyat{}|\Yat{-1}=\yat{-1};\curvpar)\propto\myexp{\natpar\cdot\genstat(\yat{}, \yat{-1})},\ \yat{},\yat{-1}\in\netsY,\label{eq:dtergm}
\end{equation}
with $\natpar=(\natparF,\natparD)$ and
  $\genstat(\yat{},\yat{-1})=(\genstatF(\yat{-1}\cup
  \yat{},\yat{-1}),\genstatD(\yat{-1}\cap \yat{},\yat{-1}))$.
\citep{KrHa14s}

\citet{KrHa14s} also noted that this formulation allowed
cross-sectional statistics developed for ERGMs to be
\enquote{converted} to a dynamic form by evaluating them on $\yF$ (so
$\genstat\form(\yF,\yat{-1})\equiv\genstat\form(\yF)$) or $\yD$ (so
$\genstat\form(\yD,\yat{-1})\equiv\genstat\form(\yD)$), which allowed them
to be interpreted in terms of tie formation and dissolution. They
called such statistics \enquote{implicitly dynamic}.

For the sake of brevity, for the rest of this paper, we will assume
that $\natpar\equiv\curvpar$, and omit $\natmap$.
\renewcommand{\natpar}{\curvpar}
\renewcommand{\natparF}{\curvpar\form}
\renewcommand{\natparD}{\curvpar\diss}
\renewcommand{\natmap}{\asdfgnm} 
\renewcommand{\Peg}{\Prob_{\genstat}}
\renewcommand{\PegF}{\Prob_{\genstat\form}}
\renewcommand{\PegD}{\Prob_{\genstat\diss}}
\renewcommand{\Eeg}{\E_{\genstat}}
\renewcommand{\Veg}{\Var_{\genstat}}
\renewcommand{\ceg}{c_{\genstat}}
\renewcommand{\cegF}{c_{\genstat\form}}
\renewcommand{\cegD}{c_{\genstat\diss}}
\renewcommand{\Oeg}{\Odds_{\genstat}}

\section{\label{sec:equil}Long-run behavior of STERGMs}
The focus of \citet{KrHa14s} was on modeling short series of
networks with an unambiguous beginning and end. Consider, instead, a
random network series $Y^0,Y^1,Y^2,\dots$ generated by the
above-described transition process. For any formation model, there is
a nonzero probability of adding any given set of allowed ties to the
network within one time step, and for any dissolution/preservation
model, there exists a nonzero probability of dissolving any given set
of extant ties. The two operate independently within each time
step. Therefore, there is a nonzero probability of transitioning from
any given network to any given network, so the transition process is
ergodic, the sequence $Y^0,Y^1,\dots$ converging to an equilibrium
distribution. \citep[Thm. 3.2.1, for example]{lefebvre2007asp}
Higher-order Markov variants also have these properties.

As the STERGM is a submodel of the TERGM, and a TERGM has the same
equilibrium properties but is more concise, we use its parametrization
for the purposes of the following discussion.
\subsection{\label{sec:equilibrium}General case}
As of this writing, we are not aware of any way to express in
algebraic form, or even in terms of a non-recursive integral, the pmf
of the equilibrium distribution $\Peg(\Yyat{};\curvpar)$, without
placing constraints on the model such as the temporal dyadic
independence discussed in Section~\ref{sec:equil-dyad-ind}.

At best, it may be defined recursively, by definition of a discrete
stationary distribution \citep[eq. (3.55), for example]{lefebvre2007asp}:
\begin{align}
  \Peg(\Yyat{};\curvpar)&=\sum_{\yat{-1}\in\netsY}\Peg(\Yyat{}\land\Yyat{-1};\curvpar)\notag\\
  &=\Eeg\left(\frac{\myexp{\natpar\cdot\genstat(\yat{},
        \Y)}}{\ceg(\curvpar,\Y)};\curvpar\right), \label{eq:equil-expectation}
\end{align}
with expectation taken over the stationary distribution itself.

\subsection{\label{sec:equil-dyad-ind}Dyadic independence}
\subsubsection{Types of dyadic independence}
As with cross-sectional ERGM, the tractable class of TERGMs is those
models with dyadic independence. In a cross-sectional ERGM context,
dyadic independence means simply that the states of all dyads are
conditionally independent given covariates \citep{hunter2008epf}:
\begin{align*}
  \Peg(\Yy;\curvpar)&=\prod_{\pij \in \dysY} \Peg(\Yij=\yij;\curvpar)\\
  &=\prod_{\pij \in \dysY} \frac{\myexp{\left(\natpar \cdot \changeij\genstat(\y) \right)\yij}}{1+\myexp{\natpar \cdot \changeij\genstat(\y)}},
\end{align*}
with
$\changeij\genstat(\y)=\genstat(\y\cup\{\pij\})-\genstat(\y\setsub\{\pij\})$,
a \emph{change statistic} vector of the model for dyad $\pij $,
which, by virtue of being a change statistic for the dyad $\pij $ does
not depend on $\yij$ and by virtue of dyadic independence does not
depend on any other dyads of $\y$, thus not depending on the state of
$\y$ at all. (\citet{krivitsky2011ans} and others further discuss
change statistics and their uses.)

A dynamic model adds another dimension to the notion of dyadic
independence. \citet{hanneke2010dtm} focused on models that have
dyadic independence only given the entirety of the network at the
previous time step. In other words, a dyad $\Yatij{}$ is conditionally
independent of a different dyad $\Yat{}_{i',j'}$, $\pipjp \ne\pij $,
but it may depend on its state in the
previous time step (i.e., $\Yat{-1}_{i',j'}$). Thus, while there is dyadic independence within a
time step, there is dyadic dependence over time. This dependence
structure is somewhat akin to the structure of the continuous-time
models of \citet{snijders2001ses}: what happens to each dyad at any
given point in time is independent of what happens to other dyads at
the time, but once some dyad does change, and the \enquote{clock}
advances, it may affect the evolution of other dyads. Models with this
structure are submodels of the STERGM, since they assume that all dyad
changes and non-changes within a time step are independent conditional
on $\Yat{-1}$, while STERGM only assumes that changes and non-changes
of ties present in $\Yat{-1}$ are independent of changes and
non-changes of non-ties of $\Yat{-1}$, conditional on $\Yat{-1}$,
within a time-step: there may be dependence within the set of those
dyads which had ties in $\Yat{-1}$ and dependence within the set of
those dyads which did not have ties in $\Yat{-1}$.

However, this within-time-step dyadic independence restriction, as of
this writing, does not appear to be sufficient to derive a closed form
for the stationary distribution. (We note that if it were, a more
general result for the stationary distribution of general
continuous-time Markov network models would have likely been available
as well.)

A further constraint, \emph{temporal dyadic independence}, that
$\Yatij{}$, may not depend on $\Yat{-1}_{i',j'}$ either, or
\begin{align*}
  \Peg(\Yyat{}|\Yyat{-1};\curvpar)&=\prod_{\pij \in \dysY} \Peg(\Yyatij{}|\Yyatij{-1};\curvpar)\\
  &=\prod_{\pij \in \dysY} \frac{\myexp{\left(\natpar \cdot \changeij\genstat(\yat{},\yat{-1}) \right)\yij}}{1+\myexp{\natpar \cdot \changeij\genstat(\yat{},\yat{-1})}},
\end{align*}
can be imposed. Then,
$\changeij\genstat(\yat{},\yat{-1})=\changeij\genstat(\yatij{-1})$ ---
the change statistic only depends on the state of the same dyad during
the previous time step. For brevity, let
$\genstat^0\sij=\changeij\genstat(\yat{},\yat{-1})|_{\yatij{-1}=0}$
and
$\genstat^1\sij=\changeij\genstat(\yat{},\yat{-1})|_{\yatij{-1}=1}$.

\subsubsection{Stationary distribution}

Under this constraint, each dyad evolves independently, forming a
2-state time-homogeneous Markov chain, which has a stationary
distribution 
with
\begin{align*}
  \Peg(\Yij=1;\curvpar)&=\frac{\Peg(\Yatij{}=1|\Yatij{-1}=0;\curvpar)}{\Peg(\Yatij{}=1|\Yatij{-1}=0;\curvpar)+\Peg(\Yatij{}=0|\Yatij{-1}=1;\curvpar)},
\end{align*}
giving
\begin{align}
  \Oeg(\Yij=1;\curvpar)
  &=\frac{1+\myexp{\natpar \cdot  \genstat^1\sij}}{1+\myexp{-\natpar \cdot  \genstat^0\sij}}\label{eq:dyad-ind-equil-odds}
\end{align}
and 
\begin{align*}
  \Peg(\Yyij;\curvpar)
  &=\prod_{\pij \in \dysY}\left(1+\frac{1+\myexp{\natpar \cdot \genstat^1\sij}}{1+\myexp{-\natpar \cdot \genstat^0\sij}}\right)^{-1}\prod_{\pij \in \dysY}\left(\frac{1+\myexp{\natpar \cdot \genstat^1\sij}}{1+\myexp{-\natpar \cdot \genstat^0\sij}}\right)^{\yij}.
\end{align*}

\subsubsection{\label{sec:ex-fed-ded}Example: Formation edge count and dissolution edge count}
Consider a STERGM with an edge count statistic for formation and edge
count statistic for dissolution/preservation. That is,
$\genstat\form(\yF,\yat{-1})=(\abs{\yF})$ and
$\genstat\diss(\yD,\yat{-1})=(\abs{\yD})$, equivalent to a TERGM with
$\genstat(\yat{},\yat{-1})=\left(\abs{\yat{}\cup\yat{-1}},\abs{\yat{}\cap\yat{-1}}\right)$,
with transition probability
\[\Peg(\Yyat{}|\Yyat{-1};\curvpar)=\frac{\myexp{\curvparF \abs{\yat{}\cup\yat{-1}}+\curvparD \abs{\yat{}\cap\yat{-1}}}}{\ceg(\curvpar,\yat{-1})}\] 
and change statistic 
$\genstat^0\sij=(+1,0)$ and $\genstat^1\sij=(0,+1)$. Substituting into~\eqref{eq:dyad-ind-equil-odds} gives an equilibrium network density
\begin{align*}\Peg(\Yat{}=1;\curvpar)
  &=\frac{1+\myexp{\curvparD}}{2+\myexp{-\curvparF}+ \myexp{\curvparD}}.
\end{align*}
In a sense, the model balances itself in the long run: having
greater-than-equilibrium number of ties gives more room for
dissolution to work, while having fewer-than-equilibrium number give
it less room and gives more room to formation.

The probability of a given tie being preserved during each time step
is simply $\ilogit(\curvparD)$. This means that the duration
distribution of a tie is simply $\Geometric(\ilogit(-\curvparD))$
(with support being $\NN$).

Notably, this is not necessarily the same as the distribution of the
time elapsed since the tie was formed as of the time of observation
(the \enquote{age} of the tie), given that the tie was observed. On
one hand, the probability that a tie is observed is proportional to
its ultimate duration, but on the other hand, every extant tie in an
observed network has its duration right-censored.  Let $X$ be the
duration a tie in the network process given that it was formed. Then,
$X_\text{obs}$, the duration of a tie given that it was observed has
the pmf
\[
  f_{X_\text{obs}}(x;\curvparD) = \frac{xf_{X}(x;\curvparD)}{\sum_{x'=1}^\infty x' f_{X}(x';\curvparD)} = \frac{xf_{X}(x;\curvparD)}{\E(X;\curvparD)}
\]
Assuming no dependence between time of observation and presence of
ties, an observed tie's duration will be right-censored at a uniform
time point. The observed age (right-censoring point)
$X_{\age|\text{obs}} \sim \Uniform(\{1,..,X_{\text{obs}}\})$, giving
it the pmf 
\begin{align*}
  f_{X_{\age|\text{obs}}}(x;\curvparD)&= \sum_{k=x}^\infty f_{X_\text{obs}}(k;\curvparD) f_{\Uniform(\{1,..,k\})}(x)\\
  &= \sum_{k=x}^\infty \frac{kf_{X}(k;\curvparD)}{\E(X;\curvparD)} \frac{1}{k}\\
  &= \frac{1-F_{X}(x-1;\curvparD)}{\E(X;\curvparD)}.
\end{align*}

For the simple case above, this distribution can be derived
in closed form: if $X\sim \Geometric(p)$,
\begin{equation}
  f_{X_{\age|\text{obs}}}(x)=\frac{(1-p)^{x-1}}{1/p}=(1-p)^{x-1}p=f_{\Geometric(p)}(x)\label{eq:geom-dur-gmme}
\end{equation}
Thus, somewhat surprisingly, the selection effect and the
right-censoring cancel exactly for geometrically-distributed
durations.

\section{\label{sec:est}Estimation based on cross-sectional data with duration information}

Momentary or cross-sectional network data observed, whether as a dyad
census or egocentrically, are a product of some social process
occurring over time. Similarly, the stationary distribution of a
stochastic process for network evolution, such as a TERGM or a
continuous-time Markov process, is a product of this evolution
process. Thus, to the extent that the model for network evolution is
an accurate model of the social process, the observed network data may
be viewed as a draw from the equilibrium distribution under the model.

In this section, we discuss avenues of estimation and inference on
STERGMs and TERGMs in general under the assumption that either the
social network evolution process (or, at least, its endogenous
components) is, or has been, fairly homogeneous (in that the
network-to-network transition probabilities do not change) over a long
time or that inhomogeneity in this process has been modeled, and thus
the social process has converged to a sort of an equilibrium. This is
a very strong assumption, which conditional approaches of
\citet{snijders2001ses}, \citet{hanneke2010dtm}, and
\citet{KrHa14s} do not require. However, these conditional
approaches require observing the network at at least two time
points. Furthermore, as we show in
Section~\ref{sec:prob-conditioning}, even if an egocentric sample is
taken at several time points, conditional methods cannot make use of
such data, without making their own, fairly strong assumptions.

\subsection{\label{sec:emle-est}Likelihood-based estimation}
Unconditional, or equilibrium, MLE for $\curvpar$ based on networks
observed at $T$ time points,
\begin{subequations}
\begin{align}\hat{\curvpar}&=\argmax_{\curvpar}\Peg(\Yyat{},\dotsc,\Yyat{+T-1};\curvpar)\notag\\
  &=\argmax_{\curvpar}\Peg(\Yyat{};\curvpar)\times\label{eq:emle-1}\\
  &\phantom{=\argmax_{\curvpar}}\Peg(\Yyat{+1},\dotsc,\Yyat{+T-1}|\Yyat{};\curvpar),\label{eq:emle-rest}\end{align}
\end{subequations}
while it depends on the equilibrium assumption, is likely to make
fuller use of the information in the data than CMLE: CMLE explicitly
ignores information, embodied in~\eqref{eq:emle-1}, about the network
evolution process that had led up to $\Yat{}$, while equilibrium MLE
would maximize the product of both~\eqref{eq:emle-rest} and~\eqref{eq:emle-1}.

A joint likelihood for an observed network series can be evaluated by
multiplying the equilibrium probability of the first observed network
by the transition probabilities of the subsequent networks, which can
be evaluated per \citet{KrHa14s}. The main difficulty with
finding the equilibrium MLE is, thus, evaluating~\eqref{eq:emle-1} (or
a ratio between them for two different values of $\curvpar$).

Derived from the definition of a stationary
distribution,~\eqref{eq:equil-expectation} suggests that it is
possible to evaluate the likelihood by simulation: the probability of
observing a network $\yat{}$ is the expectation over the possible
networks $\Yat{-1}$ drawn from the stationary distribution that could
have transitioned to $\yat{}$, of the probability of transitioning
from that $\Yat{-1}$ to $\yat{}$. In this case,
$\ceg(\curvpar,\Yat{-1})$ depends on $\Yat{-1}$ and is inside the
expectation, so it is not constant.  It is fairly straightforward to
simulate from the stationary distribution (it being the equilibrium
distribution), and evaluation of the transition probability, though
imprecise, may be possible, thus allowing the likelihood to be
approximated. This estimation is likely to further suffer from the
problem that for almost any realistic dynamic network process, the
probability of transitioning from vast majority of random equilibrium
draws $\Y$ to the observed $\yat{}$ will be very small. This is
because a typical network is expected to change slowly over time, so
only $\Y$ which are similar to $\yat{}$ significantly contribute to
the likelihood. This is not very different from the problem of
evaluating Bayes factors using only direct simulation.

Furthermore, if only a single network is observed, even if the
equilibrium likelihood could be could be evaluated, it is unlikely to
be sufficient for estimating a dynamic model. For example, in the
model in Section~\ref{sec:ex-fed-ded}, $\abs{\y}$ is sufficient for
both $\curvparD$ and $\curvparF$: two parameters with only one
sufficient statistic, leading to nonidentifiability: one cannot infer both tie incidence and tie duration from tie prevalence alone \citep{KrHa14s}. Thus, to fit
these models, some form of temporal information is required. This
temporal information may be a part of the dataset, in the form of
\enquote{ages} of extant ties, or it may come from a different
source. For data of interest --- networks of sexual partnerships ---
one survey may contain information about momentary network behavior
(e.g., number of ongoing sexual partnerships of respondent has at the
time of the survey) while a different survey with respondents drawn
from similar population may contain information about partnership
duration distribution.

Making use of such information using maximum likelihood estimation
requires integrating over possible network series drawn from the
equilibrium distribution to derive the equilibrium distribution at
$\curvpar$ of available statistics. This appears to be infeasible at
this time. However, simulating from a network series is
straightforward, so we turn to generalized method of moments
estimation instead.

\subsection{Generalized Method of Moments Estimation}
A Generalized Method of Moments Estimator (GMME) seeks that parameter
configuration $\tilde{\curvpar}$ such that the expected value of the
statistic of interest matches its observed value. That is, for some
network process defined by $\genstat$ and parametrized by $\curvpar$,
let $\target(\yat{},\yat{-1},\dotsc)$ (\emph{targeted} statistic) be a
vector function of a network or network series, whose values have
been observed or inferred from available data, and let
\begin{equation}\meanstats(\curvpar)\equiv\Eeg\left(\target(\Yat{+1},\dots,\Yat{+T});\curvpar\right),\label{eq:gmme-obj-mean}\end{equation}
its expected value under the network process of interest, and
\begin{equation}\V(\curvpar)\equiv\Veg\left(\target(\Yat{+1},\dots,\Yat{+T});\curvpar\right)\label{eq:gmme-obj-var}\end{equation}
its variance-covariance matrix. Then, for some observed statistic of
interest $\target(\y^1,\dots,\y^T)$ an optimal GMME minimizes an
objective function
\begin{equation}
  J(\curvpar)\equiv\left(\meanstats(\curvpar)-\target(\y^1,\dots,\y^T)\right)\t \V(\curvpar)^{-1} \left(\meanstats(\curvpar)-\target(\y^1,\dots,\y^T)\right),\label{eq:gmme-obj}
\end{equation}
the squared Mahalanobis distance between the observed value of $\target$ and
its expected value. \citep{hansen1996fsp} (If the
dimension of $\curvpar$ is the same as the dimension of $\target$,
often, at GMME $\tilde\curvpar$, $J(\tilde\curvpar)=0$.)  This
estimator has asymptotic variance
$\Veg(\tilde\curvpar)=(\gradient(\tilde\curvpar)\t\V(\tilde\curvpar)^{-1}\gradient(\tilde\curvpar))^{-1}$,
where $\gradient(\curvpar)=\partial \meanstats(\curvpar) /\partial
\curvpar$, the gradient matrix of the mapping from the model
parameters to the expected values of target statistics, although
asymptotic results applied to network models fit to a single network
are tenuous \citep{hunter2006ice}. In linear exponential families
(including linear ERGMs), and with $\genstat(\y)\equiv\target(\y)$,
MLE and GMME are the same \citep[p. 367--368]{casella2002si}, but we
may want to separate the two, because the sufficient statistic may not
have been observed as is the case in the previous sections, while GMME
can make use of any statistics available.

To distinguish it from the targeted statistic, we refer to $\genstat$
as the \emph{generative} statistic. Whereas the elements of the
generative statistic are chosen based on beliefs about the nature of
the social process being modeled, targeted statistic's elements are
determined by what data about the network or network series of
interest are available and what statistics are likely to be
informative about the generative model's free
parameters. 
We illustrate this distinction in a cross-sectional context.

\subsubsection{Example: Edge count generative statistic with an isolate count target}
Suppose that an undirected network of $n$ actors
is modeled as an \ErRe{} graph, that is,
\[\Peg(\Yy;\curvpar)=\frac{\myexp{\curvpar\cdot \abs{\y}}}{\ceg(\curvpar)}, \qquad\ceg(\curvpar)=(1+\myexp{\curvpar})^{\abs{\dysY}},\]
but due to the nature of the observation process, only the
$\target(\y)=\sum_{\iactors} \I{\abs{\y_i}=0}$, the number of isolates
in the network, has been observed, while the sufficient
statistic is $\genstat(\y)=\abs{\y}$.

The MLE for $\curvpar$ given data available,
\begin{align*}
  \hat{\curvpar}&=\argmax_\curvpar\Peg(\target(\Y)=\target(\y);\curvpar)\\
  &=\argmax_\curvpar \sum_{\y'\in \netsY}\I{\target(\y)=\sum_{\iactors} \I{\lvert \
      y'_i\rvert=0}}\frac{\myexp{\curvpar\cdot \abs{\y}}}{\ceg(\curvpar)}.
\end{align*}
This probability is not straightforward to evaluate on an undirected
network. On the other hand, it is straightforward to evaluate
\begin{align*}
  \meanstats(\curvpar)&=\Eeg\left(\sum_{\iactors} \I{\abs{\Y_i}=0};\curvpar\right)\\
  &=\sum_{\iactors} \Peg(\abs{\Y_i}=0;\curvpar)\\
  &=n \left(1-\ilogit(\curvpar)\right)^{n-1},
\end{align*}
so GMME
\begin{align*}
  \tilde{\curvpar}&=\curvpar:\Eeg(\target(\Y);\curvpar)=\target(\y)\\
  &=\logit\left(1-\left(\frac{\target(\y)}{n}\right)^{\frac{1}{n-1}}\right).
\end{align*}
From the point of view of GMME, any correspondence between the
generative and the targeted statistic is entirely arbitrary.  

To emphasize our focus on using target statistics evaluated on the
equilibrium distribution induced by a dynamic model, we refer to such
an estimate as the \emph{Equilibrium Generalized Method of Moments
  Estimator} (EGMME).

\subsection{\label{sec:target-stats}Selecting target statistics}
In CMLE, the sufficient statistic for the model parameters is simply
the $\genstat(\yat{},\yat{-1})$, the generative statistic. When only a
single network is observed, even with durational information, the
generative statistic may not be possible to evaluate on the data, so
the only hard requirement is that $\target$ must contain information
about $\genstat$. In this section, we briefly address the question of
how $\target$ might be selected. (\citet{snijders2007mcn} dealt with a
similar problem, albeit conditioning on the initial network.) Notably,
in GMME, the dimension of $\target$ may exceed the dimension of
$\genstat$, which means that if there are multiple candidate targets,
it may be practical to simply use them all.

\subsubsection{Target statistics for structure}
If a generative statistic $\genstatF_k$ or $\genstatD_k$ is implicitly
dynamic ($\genstatF_k(\yF,\yat{-1})\equiv\genstatF_k(\yF)$
and similarly for $\genstatD_k(\yD,\yat{-1})$) it is likely, though
not certain, that increasing their coefficient will increase
$\Eeg(\genstatF_k(\Yat{}))$ or $\Eeg(\genstatD_k(\Yat{}))$,
respectively. For example, if $\genstatF_k(\yF)=\abs{\yF}$, the edge
count, increasing the coefficient on $\genstatF_k(\yF)$ will increase
the expected number of edges at the equilibrium. This suggests, though
does not prove, that when implicitly dynamic statistics are used,
their cross-sectional progenitors may make near-optimal targets.

\subsubsection{Target statistics for duration}
A network cross-section alone is not sufficient to estimate a model
with free parameters in both formation and dissolution. One form of
duration information comprises duration of past ties (often within
some fixed and known time window) and \enquote{ages} of ongoing
ties. For example, in the NHSLS, the respondents were asked to
enumerate their sexual partnerships in the 12 months preceding the
interview \citep{laumann1994sos}, and, for each partnership, were
asked how many months before the interview it had started and how many
months before the interview it had ended.

From the point of view of survival analysis, this makes the data
right-censored and left-truncated
\citep[pp. 72--78]{klein2003sat}. Estimates of hazard structure from
survival analysis could then be used to specify the dissolution model,
fitting formation model conditional on that.

A more direct EGMME-based approach is to use duration-sensitive
statistics, such as the average age of an extant tie, as targets. In
some cases, with a simple dissolution model (and an arbitrarily
complex formation model), it may be possible to estimate dissolution
in closed form (e.g., \eqref{eq:geom-dur-gmme}). Otherwise, to the
extent that the assumption that the observed process is an equilibrium
draw is valid, no special adjustment for censoring or truncation is
needed: the duration of a simulated edge existing at a given time will
be censored (since it still exists) and left-truncated (since if it
had dissolved earlier, it would not have been observed) just like the
data, so matching the expected value for this quantity to the observed
will arrive at the corrected estimate.

\section{\label{sec:est-ego}Estimation based on egocentrically sampled data}
Per the discussion in the Introduction, data on sexual partnership
networks are rarely available in the form of a dyad census, much less
a series of network observations over time or exact timings of tie
formation and dissolution. It is much more typical to observe an
egocentric sample from the population of interest. Per
\citet{krivitsky2011ans}, let $E$ be the set of respondents in an
egocentric survey (\enquote{egos}), and, for each ego $e\in E$, let
$A_e$ be the set of nominations (\enquote{alters}) of $e$, and let
$A=\bigcup_{e\in E}A_e$. While each $e\in E$ represents a distinct
actor in the network of interest, $a\in A$ represent
\emph{nominations}, not actors: multiple respondents may (unknowingly)
nominate the same actor or may nominate each other. Each actor (ego
and alter) is associated with a set of attributes, denoted $\covariate_e$
and $\covariate_a$.

\citet{krivitsky2011ans} fit cross-sectional ERGMs to egocentrically
sampled data by constructing a model invariant to network size, in
that for two networks of different size but similar features of
interest (mean degree, degree distribution, and selective mixing) the
model produce (asymptotically) similar MLEs, and, conversely, a given
parameter configuration would induce distributions of networks with
similar features of interest across a variety of network sizes and
compositions. They then considered a hypothetical network made up of
the respondents in the survey and having similar structure as the
population network, and computed the statistic of interest that that
network would have to have in order to have produced the observed
egocentric sample as an egocentric census; and this statistic was used
to fit an ERGM.  More concretely, in an undirected network, where each
tie $\pij$ has the potential to be reported twice --- once by $i$ and
once by $j$, a dyad-level statistic of the form
$\genstat_\covariate(\y)=\sum_{\ijdysY} \yij
f(\covariate_i,\covariate_j)$, for some function $f$ of the attributes
$\covariate$ of $i$ and $j$, could be recovered from an egocentric
census as $\half\sum_{e\in E} \sum_{a\in A_e}
f(\covariate_e,\covariate_a)$, and for an actor-level statistic of the
form $\genstat_\covariate(\y)=\sum_{\iactors} f(\covariate_i,
\covariate_j: j\in \y_i)$, for $f$ a function of the attributes of $i$
and attributes of $i$'s neighbors, including their number, as
$\sum_{e\in E} f(\covariate_e,\covariate_a: a\in A_e)$.

Here, we extend this approach to fit dynamic ERGMs to egocentrically
sampled data. Let $A_e^t$ be the egocentric observation of alters
network at time $t$ (the list of respondents $E$ assumed to be
unchanging), and similarly to the earlier sections, define the
\emph{age} of a nomination, $\age(a)$ to be amount of time elapsed
since the relationship (ongoing at time $t$) began. In egocentric
surveys of sexual history, relationships ongoing at the time $t$ of
the survey ($A^t$) are most reliably observed, but many surveys such as the
National Health and Social Life Survey (NHSLS) \citep{laumann1994sos}
also ask the respondents to enumerate all relationships (and their
durations) that had ended in the prior $T$ time units. Thus, $A^{t-T},
\dotsc, A^{t-1}$ may also observed and, furthermore it is possible to
distinguish between alters incident on the same ego at different time
points.

In the following, we derive what is needed to fit a Conditional MLE
(CMLE) from egocentrically sampled data, and show that such data do
not suffice except for very simple models.

\subsection{Conditional MLE}
\subsubsection{\label{sec:cmle-stats}Conditional MLE statistics}

Generally, a single network observation, even if it contains the age
of each tie, and, in particular, and egocentric observation
$(E,A^t,\covariate^t)$ at a single time point, is not sufficient to
evaluate the CMLE for the TERGM (separable or not) for a
transition. To see why, consider a single dyad across two time points:
$(\yatij{-1},\yatij{})$. It has four possible histories: $(0,0)$ (no
tie), $(0,1)$ (tie formed), $(1,0)$ (tie dissolved), and $(1,1)$ (tie
preserved). If only $\yatij{}$ is observed, it is possible to identify
histories $(0,1)$ and $(1,1)$ from $(0,0)$ and $(1,0)$, and if
$\age(\yatij{})$ is also observed, it is further possible to identify
$(0,1)$ from $(1,1)$, with $\age(\yatij{})\le 1$ indicating the former
and $\age(\yatij{})>1$ indicating the latter, but it is not possible
to identify $(0,0)$ from $(1,0)$ from $\yatij{}$ and $\age(\yatij{})$
alone. In the context of TERGM, terms like stability (the number of
dyads whose state has changed) cannot be evaluated; and in STERGMs in
particular, no implicitly dynamic dissolution terms can be (though
formation terms can).

If egocentric observations at multiple time points are available, the
sufficient statistic for many transition models can be evaluated. For
example, a statistic of the form
\begin{multline*}\genstat_\covariate(\yat{},\yat{-1}) =\\ \sum_{\pij \in \yatij{-1}\cup\yatij{}}\left( (1-\yatij{-1})\yatij{} f\form(\covariate_i,\covariate_j) + \yatij{-1}(1-\yatij{}) f\diss(\covariate_i,\covariate_j)  + \yatij{-1}\yatij{} f(\covariate_i,\covariate_j) \right):\end{multline*}
effectively a sum over those dyads that can be observed of some
function of the incident actors' attributes that depends on whether a
tie was added, removed, or persisted, can be recovered as
\[\half \sum_{e\in E} \left(\sum_{a\in A_e^t\setsub A_e^{t-1}} f\form(\covariate_e,\covariate_a)+\sum_{a\in A_e^{t-1} \setsub A_e^t} f\diss(\covariate_e,\covariate_a)+\sum_{a\in  A_e^{t-1}\cap A_e^t} f(\covariate_e,\covariate_a)\right)\]
in an undirected network (or a directed network with observed
in-ties). This statistic has a variety of dyad-independent implicitly
dynamic statistics as special cases.

Similarly, actor-level statistics of the form
\[
  g_k(\yat{},\yat{-1}) = \sum_{\iactors} f(\covariate_i,
\covariate_j: j\in \yat{-1}_i, \covariate_j: j\in \yat{}_i)
\]
where $f$ is local per \citet{krivitsky2011ans}, in that it only
depends on exogenous attributes of actor $i$ and actors $j\in
\yat{}_i\cup\yat{-1}_i$ could be recovered as $\sum_{e\in E}
f(e,\covariate_a: a\in A^{t-1}_e, \covariate_a: a\in A^t_i)$.  These
include statistics such as the number of monogamous ties, including
their Inherited variants $\genstat(\yat{-1},\yat{})=\sum_{\iactors}
1_{\abs{\yat{-1}\cup\yat{}}=1}$ and
$\genstat(\yat{-1},\yat{})=\sum_{\iactors}
1_{\abs{\yat{-1}\cap\yat{}}=1}$.

\subsubsection{\label{sec:prob-conditioning}The problem of conditioning}
Section~\ref{sec:cmle-stats} described how the sufficient statistic of
the transition between two networks can be computed based on
egocentrically sampled data, but they are not, in general, sufficient
for evaluation of the likelihood, since the likelihood for the
transition,
\[
L(\curvpar)=\frac{\myexp{\natpar\cdot\genstat(\yat{}, \yat{-1})}}{\sum_{\ypnetsY} \myexp{\natpar\cdot\genstat(\y',\yat{-1})}}, 
\]
has a normalizing constant that can depend on the prior network $\yat{-1}$, even if $\genstat(\yat{}, \yat{-1})$ can be inferred from the
data.

For simple models, it may be possible to evaluate it anyway. For
example, for a STERGM described in Section~\ref{sec:ex-fed-ded},
\begin{align*}
  \ceg(\curvpar)
  &=\sum_{\ynetsY} \myexp{\curvparF\abs{\yat{-1}\cup\y}+\curvparD\abs{\yat{-1}\cap\y}}\\
  &=\left( 1 + \myexp{\curvparF}\right)^{\abs{\dysY\setsub\yatij{-1}}}\left( \myexp{\curvparF} + \myexp{\curvparF+\curvparD}\right)^{\abs{\yatij{-1}}},
\end{align*}
and $\abs{\yat{-1}}$ can be evaluated given $(E,A^{t-1}_e)$. 

\begin{figure}
  \noindent\begin{center}
    \begin{tabular}{ccc}
      \includegraphics[width=0.3\columnwidth,keepaspectratio]{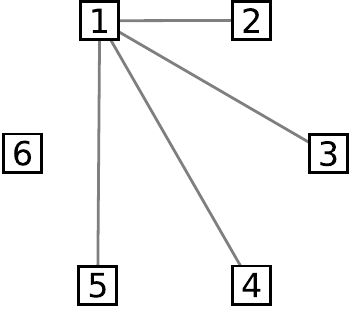}&\phantom{Not best way}
      &
      \includegraphics[width=0.3\columnwidth,keepaspectratio]{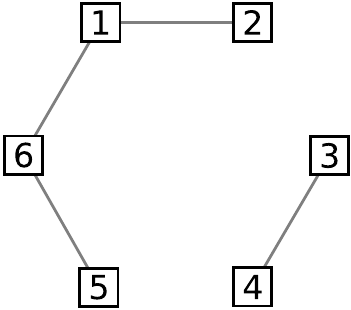}\\
      &&\\
      (a) & &  (b)  
    \end{tabular}
  \end{center}
  \caption[Networks with same statistics but different TERGM
  normalizing constants]{\label{fig:cond-recover} Both networks
    have 4 ties and 4 actors with degree 1. However, in network (b), it
    is possible to toggle one dyad ($\pij =(1,6)$) to create a network with 6
    actors with degree 1, something not possible in network (a).}
\end{figure}

However, it may not be possible to infer normalizing constants in this
way for more complex models. Consider a model with transition
statistic
\[\genstat(\y,\y')=\left(\abs{\y\Delta \y'},\sum_{\iactors}\I{\abs{\y_i}=1}\right):\]
Hamming distance (i.e. number of changed dyads) between $\y$ and $\y'$
and the number of actors in $\y$ with degree 1. Suppose that the
relevant statistic inferred from egocentric data is
\[\genstat(\yat{-1})=\left(\abs{\yat{-1}},\sum_{\iactors} \I{\abs{\yat{-1}_i}=1}\right)=(4,4),\]
the number of ties in $\yat{-1}$ and number of actors in $\yat{-1}$
with degree 1. But, while either network in
Figure~\ref{fig:cond-recover} has $\genstat^0(\yat{-1})=(4,4)$, the
normalizing constant conditional on network in
Figure~\ref{fig:cond-recover}(b) would contain a summand
$\myexp{\natpar\cdot(1,6)}$, since toggling 1 tie in that network can
result in a network with 6 actors with degree 1, while no such 1-tie
change exists for the network in Figure~\ref{fig:cond-recover}(a), and
$\myexp{\natpar\cdot(1,6)}$ would not be a summand in the normalizing
constant conditional on it. Thus, the normalizing constants would be different.

It may be possible to circumvent this problem. If an ERGM is fit to
$\yat{-1}$'s statistic, as \citet{hanneke2010dtm} suggest in a
slightly different context, it may be possible to approximate the
likelihood by integrating over $\Yat{-1}$ drawn from the induced ERGM
distribution. This may be difficult for reasons similar to those
discussed in Section~\ref{sec:emle-est}.

Another disadvantage of CMLE in this context is that it might not make
use of all available duration information. The NHSLS survey, for
instance, only asked about relationships that ended in the 12 months
prior to the survey (and asked for relationship length in
months). Even if the problem of conditioning could be circumvented,
this means that, at least using the inference of
Section~\ref{sec:cmle-stats}, the transition statistic could only be
recovered for twelve distinct time points. Any relationships that
lasted longer than that --- which is most of them --- would contribute
no additional information. For this reason, we advocate EGMME for
these problems.

\section{\label{sec:dyn-netsize}Adjusting the STERGM network evolution
  process for network size}
So far, we have treated the network size and
composition as fixed, and focused only on the evolution of network
ties. In practice, many dynamic network applications involve actors
entering and leaving the network and individual actor attributes
changing over time. In this section, we incorporate the adjustments
proposed by \citet{krivitsky2011ans} into the dynamic models, and
explore the properties of the resulting process.

The offset described by \citet{krivitsky2011ans} operates as a
coefficient on an edge count statistic. As \citet{KrHa14s}
discuss, the edge count statistic in a TERGM transition probability
has a dual effect. In particular, an increasingly negative coefficient
of the offset term of a growing network would both reduce the
incidence and shorten duration. Whether or not this is a desirable
property of the model depends on the network process of interest, but
in particular, in human sexual relationship networks, there is little
reason to believe that duration would change significantly as the
population grows or shrinks. Thus, the offset should only affect the
incidence of ties. In a separable parametrization, this is easily
achieved by only be adding the offset term to the formation phase of
the process. In other words,
\begin{multline*}\PegF(\Yya{+}|\Yyat{-1};\curvparF )=\\\frac{\myexp{\log\left(\frac{1}{n^{(t-1)}}\right) \abs{\yF} + \natparF\cdot \genstat\form(\yF,\yat{-1})}\I{\yF\supseteq \yat{-1}}}{\cegF(\curvparF ,\yat{-1})}\end{multline*}
for $n^{(t-1)}$ being the number of actors in the network at time
$t-1$; and $\PegD(\Yya{-}|\Yyat{-1};\curvparD)$ is unchanged.

In particular consider adding a network size offset to the example Section~\ref{sec:ex-fed-ded}, with edge count formation and edge count dissolution. Adding the offset to formation yields the transition
probability:
\begin{multline*}\Peg(\Yyat{}|\Yyat{-1};\curvpar)=\\\frac{\myexp{\left(\curvparF -\log(n)\right)\abs{\yat{}\cup\yat{-1}}+\curvparD \abs{\yat{}\cap\yat{-1}}}}{\ceg(\curvpar,\yat{-1})},\end{multline*}
which, substituting into~\eqref{eq:dyad-ind-equil-odds}, and
converting odds to probability gives mean degree, for an undirected
network, of
\begin{align*}
  (n-1)\Peg(\Yij=1;\curvpar)
  &=\frac{(n-1)(1+\myexp{\curvparD})}{1+n\myexp{-\curvparF}+1+\myexp{\curvparD}},
\end{align*}
which asymptotically converges to
\begin{align*}
  \lim_{n\to\infty}(n-1)\Peg(\Yij=1;\curvpar)&=\lim_{n\to\infty}\frac{(n-1)(1+\myexp{\curvparD})}{1+n\myexp{-\curvparF}+1+\myexp{\curvparD}}\\
  &=\myexp{\curvparF}+\myexp{\curvparF +\curvparD}.
\end{align*}
Thus, the offset term can be used to stabilize and control a dynamic
model's equilibrium mean degree as well.

\section{\label{sec:example}Application to dynamic population simulation based on the National Health and Social Life Survey data}
We demonstrate these ideas by taking the National Health and Social
Life Survey (NHSLS) data and model fit by \citet{krivitsky2011ans},
fitting a dynamic version of the model --- easily converted from the
cross-sectional using implicitly dynamic statistics, and then
simulating an evolving and growing population based on the estimated
parameters.  The EGMME approach is particularly well-suited so
problems where the ultimate goal is simulation: in a successful fit,
for which $J(\tilde\curvpar)=0$ will, by construction, produce a
simulation whose expected values of statistics of interest will match
those observed exactly.

\subsection{Model for network evolution}
Our exploratory survival analysis showed that a geometric distribution
was an adequate approximation to the duration distribution of a
relationship, and that there was little mixing structure in the
dissolution hazard. Therefore, we postulate an approximately geometric
relationship duration distribution, and set
$\genstatD(\yD,\yat{-1})=\abs{\yD}.$

This means that we attribute all the structure in the network to
differences in incidence of relations. Per \citet{KrHa14s},
this makes some intuitive sense: once a relationship is formed, its
persistence is likely affected by fewer factors than its
formation. Thus, we \enquote{convert} the NHSLS model of
\citet{krivitsky2011ans} to a dynamic one by only using Inherited
statistics and setting the formation statistic $\genstat\form(\yF)$ to
the same statistic as that in that article (though evaluated on $\yF$
rather than on $\y$). The terms used are listed in
Table~\ref{tab:NHSLS-ML-results} and described in more detail by
\citet{krivitsky2011ans}. We also add a network size offset of
Section~\ref{sec:dyn-netsize}.

\subsection{Inference}
We now describe the procedure for fitting this model to the available
data.
\subsubsection{Target statistic}
Because our goal is simulation, and for simplicity, we set
\[\target(\y)=\left(\genstat\form(\y),\frac{1}{\abs{\y}}\sum_{\pij \in \y} \age(\yij)\right),\]
the formation generative statistic evaluated on the cross-sectional
network (simulated or inferred) and the average age of an extant tie
in this network. This also serves to simplify fitting of this model,
because one can fairly safely assume that $\gradient(\curvpar)$ has a
positive diagonal:
$\partial\meanstats(\curvpar_k)/\partial\curvpar_k>0$.

For this demonstration, we resample the egos in the network to
resample size 1000.

\subsubsection{Time step size}
How much time is represented by a single discrete time step is a
trade-off between granularity and computational cost: shorter time
steps have a higher cost to simulate, while longer time steps make the
separability assumption that formation and dissolution are
conditionally independent within a time step less
plausible. \citep{KrHa14s} Driven in part by the format of
the data, where all duration measurements are integral counts of
months, we set
\[1\text{ time step}=1\text{ month}=\frac{1}{12}\text{ year}.\]

\subsubsection{Time-varying exogenous covariates}
{A network series simulated under $\curvpar$ of interest
  converges to equilibrium, and statistics of that chain can be used
  as input to the search for EGMME. However, the model used includes
  effects of actors' ages and age differences on relationship
  incidence, and actors age over time. When fitting model parameters,
  we ignore this: even as the network evolution process runs forward,
  all actors' ages stay the same. Nor are actors added or removed from
  the network. This is likely to produce biased estimates, such that
  the simulation stage, which does incorporate these vital dynamics,
  will not reproduce the statistics of interest exactly, particularly
  the age-related statistics.}
\subsection{Simulation}
{We simulate the evolution of a network of sexual
  partnerships based on the network process described above, over 500
  simulation years (6,000 1-month time steps), incorporating a model
  with vital dynamics: aging (with actors aging out of the population
  at 60, maintaining a closed network of 18--59), actors randomly
  removed from the population, and actors randomly
  \enquote{reproducing}, to test the network size adjustment.

}
\subsubsection{Population process model}
{We use a starting network constructed out of the resampled
  \enquote{egos}, with simulated annealing used to find a particular
  configuration of ties that has cross-sectional target statistic similar to
  that inferred. In the network (tie) evolution model, we assume that
  tie formation and dissolution do not affect each other within a time
  step, and our incorporation of vital dynamics is done similarly:
  within each time step the network evolution process takes place,
  then population evolution process (not affected by network evolution
  process) takes place. Thus, within a time step, the vital dynamics
  changes do not affect the network process, and the network process
  does not affect the vital dynamics changes.

  The following procedure is iterated every time step (month):
  \begin{enumerate}[partopsep=0pt,topsep=0pt,parsep=0pt,itemsep=0pt]
  \item The network evolution model, adjusted for network size, is run
    one time step forward, forming and dissolving network ties.
  \item For each actor, with probability 0.0023, an identical actor,
    but aged 18 and with sex selected at random is added to the
    population.
  \item For each actor, with probability 0.00042, the actor is removed
    from the population.
  \item Actors' ages are incremented by 1 month.
  \item For each actor, if the actor's age equals or exceeds 60, the
    actor is also removed from the population.
  \item All ties incident on removed actors are dissolved.
  \item The state of the population is recorded.
  \end{enumerate}
  The specific values for \enquote{birth rate}, \enquote{death rate}
  were selected to give a modest but substantial population growth
  over the length of the simulation.
}

\subsection{\label{sec:implementation}Implementation}
Some of the difficulties associated with fitting these models and the
algorithm we use are discussed in Appendix~\ref{app:gmme}. As a simple
shortcut, we use \eqref{eq:geom-dur-gmme} to estimate 
\[\tilde{\curvpar}=-\logit\left(\left(\frac{1}{\abs{\y}}\sum_{\pij \in \y} \age(\yij)\right)^{-1}\right).\]

Software package \pkg{ergm} \citep{hunter2008epf,handcock2012epf} in
the \pkg{statnet} \citep{handcock2008sst} suite of libraries for
social network analysis in \proglang{R}
\citep{rdevelopmentcoreteam2009rla} was used to fit and simulate from
these models, with the package \pkg{networkDynamic}
\citep{lesliecook2012nde} used to store and inspect the simulation
results.

\subsection{Results}
\subsubsection{\label{sec:NHSLS-model-fit}Model fit}
\begin{figure} \noindent {\centering
\includegraphics[width=0.95\columnwidth,keepaspectratio]{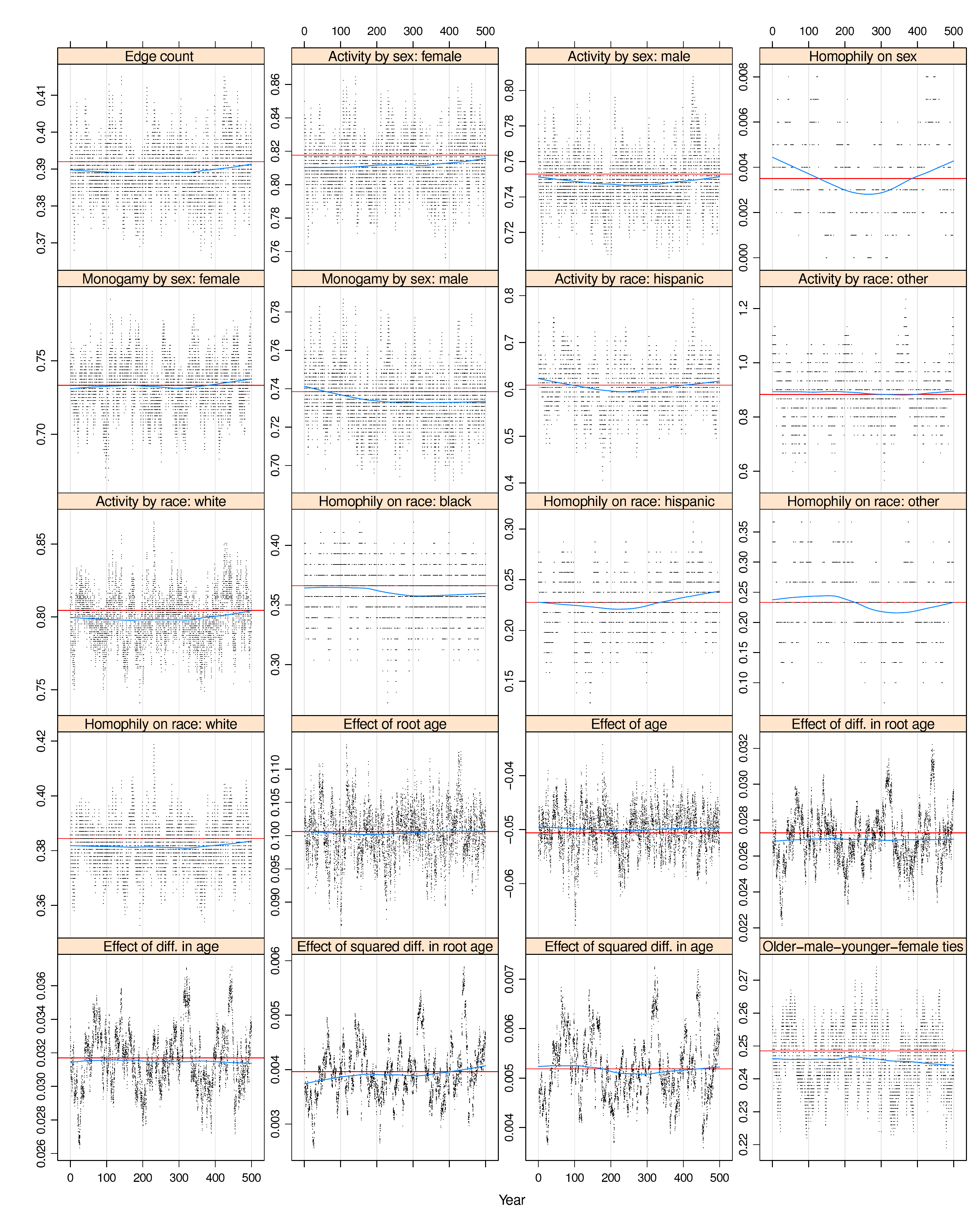}
}
\caption[Simulated network statistics in a static
population]{\label{fig:NHSLS-ML0-results} NHSLS dynamic simulation
  network statistics over time, in a static population. All values
  are normalized per Section~\ref{sec:NHSLS-model-fit}. Red lines are the \enquote{target} values for their
  respective statistics.}
\end{figure}

The inferred equilibrium target statistic and EGMME parameters are
given in Table~\ref{tab:NHSLS-ML-results}. To test the fit, we
simulated a network series from the model over a static population:
same values for $\curvpar\form$ and $\curvpar\diss$ were used and the
same initial network, but no actors were added or removed, and actor
ages were held fixed. The simulated values (shown in
Figure~\ref{fig:NHSLS-ML0-results}) are essentially the same as those
observed, suggesting that the correct fit was found. For
interpretability, we normalize network statistics we report as
follows:
\begin{enumerate}[partopsep=0pt,topsep=0pt,parsep=0pt,itemsep=0pt]
\item if a statistic pertains directly to a particular subset of
  actors, (e.g. a particular sex or race category), it is normalized by
  the total number of actors in that subset at the time;
\item otherwise, the statistic is normalized by the overall network
  size at the time.
\end{enumerate}
Thus, for example, the statistic reported for the actor activity by
sex for males is the number of ties incident on male actors (with
male-male ties counting twice) divided by the number of male actors:
that is, the mean degree of a male in the network. Similarly, the
statistic for the monogamy for females (number of female actors with
degree 1) is reported as the proportion of the total number of female
actors in the network. On the other hand, age effects do not pertain
to any particular group, and are thus normalized by the whole
network's size.

\begin{table}
\caption[STERGM parameter estimates and population simulation results]{\label{tab:NHSLS-ML-results} Parameter estimates and simulation results\\ All network statistics ($\target$) are per capita in the group they describe (per Section~\ref{sec:NHSLS-model-fit}); times are in years.}
\begin{center}
\begin{tabular}{|l|rr|rrr|}
  \hline
  &\multicolumn{2}{c|}{Inference} & \multicolumn{3}{c|}{Simulation}\\
  \hline
 &$\target(\y)$& $\tilde\curvpar$ & $\target(\Y^{100})$&$\target(\Y^{300})$ &$\target(\Y^{500})$ \\
 Network size                                             &$  1000  $&$  1000  $&$  1469  $&$  2111  $&$  3227  $\\
 \hline
 Formation & & && & \\
 \quad Offset                                             &$        $&$ -6.908 $&&& \\
 \quad Actor activity by sex &&& & & \\
 \quad \quad Female                                       &$  0.819 $&$ -4.563 $&$  0.713 $&$  0.740 $&$  0.715 $\\
 \quad \quad Male                                         &$  0.756 $&$ -5.520 $&$  0.717 $&$  0.713 $&$  0.701 $\\
 \quad Same-sex partnership                               &$  0.004 $&$ -4.318 $&$  0.005 $&$  0.004 $&$  0.004 $\\
 \quad Monogamy by sex &&&& & \\
 \quad \quad Female                                       &$  0.735 $&$  2.007 $&$  0.637 $&$  0.694 $&$  0.658 $\\
 \quad \quad Male                                         &$  0.740 $&$  2.793 $&$  0.717 $&$  0.701 $&$  0.692 $\\
 \quad Actor activity by race &&&& & \\
 \quad \quad Black                                        &  (base)  &  (base)  &          &          &          \\
 \quad \quad Hispanic                                     &$  0.614 $&$  0.770 $&$  0.504 $&$  0.478 $&$  0.414 $\\
 \quad \quad Other                                        &$  0.900 $&$  2.605 $&$  0.820 $&$  0.898 $&$  0.905 $\\
 \quad \quad White                                        &$  0.806 $&$  1.175 $&$  0.726 $&$  0.735 $&$  0.710 $\\
 \quad Race homophily by race &&& & & \\
 \quad \quad Black                                        &$  0.366 $&$  6.334 $&$  0.347 $&$  0.316 $&$  0.341 $\\
 \quad \quad Hispanic                                     &$  0.228 $&$  3.226 $&$  0.160 $&$  0.147 $&$  0.133 $\\
 \quad \quad Other                                        &$  0.233 $&$  2.724 $&$  0.220 $&$  0.320 $&$  0.319 $\\
 \quad \quad White                                        &$  0.384 $&$  2.017 $&$  0.344 $&$  0.349 $&$  0.337 $\\
 \quad Age effects &&&& & \\
 \quad \quad $\sqrt{\text{age}}$ effect                   &$  0.101 $&$  3.725 $&$  0.127 $&$  0.136 $&$  0.135 $\\
 \quad \quad $\text{age}$ effect                          &$ -0.051 $&$ -2.735 $&$  0.002 $&$  0.010 $&$  0.011 $\\
 \quad Age difference effects &&&&&\\
 \quad \quad Difference in $\sqrt{\text{age}}$            &$  0.027 $&$ -5.801 $&$  0.019 $&$  0.018 $&$  0.018 $\\
 \quad \quad Difference in $\text{age}$                   &$  0.032 $&$ -9.012 $&$  0.025 $&$  0.024 $&$  0.023 $\\
 \quad \quad Squared difference in $\sqrt{\text{age}}$    &$  0.004 $&$  5.234 $&$  0.002 $&$  0.002 $&$  0.002 $\\
 \quad \quad Squared difference in $\text{age}$           &$  0.005 $&$  2.264 $&$  0.004 $&$  0.003 $&$  0.003 $\\
 \quad \quad Older-male-younger-female                    &$  0.249 $&$  1.116 $&$  0.176 $&$  0.178 $&$  0.158 $\\
 \hline
 Dissolution & & & & & \\
 \quad Mean duration (years)                              &$ 10.307 $&$  4.810 $ & \multicolumn{3}{c|}{$  7.160 $ raw, $ 10.181 $ adjusted} \\
 \hline
\end{tabular}
\end{center}
\end{table}

\subsubsection{Evolution of network structure}
{Over the course of the simulation with evolving size and
  composition, the population grew from 1,000 actors to 3,227 actors
  (shown in Figure~\ref{fig:NHSLS-ML-n}(a)), with a total of 28,652 individuals
  in the population during the period. Because the population is
  fairly small and the population growth is exponential,
  there were substantial changes in population composition (shown in Figure~\ref{fig:NHSLS-ML-n}(b)).

  \begin{figure} \noindent {\begin{center}
        \begin{tabular}{cc}
          \includegraphics[width=0.48\columnwidth,keepaspectratio]{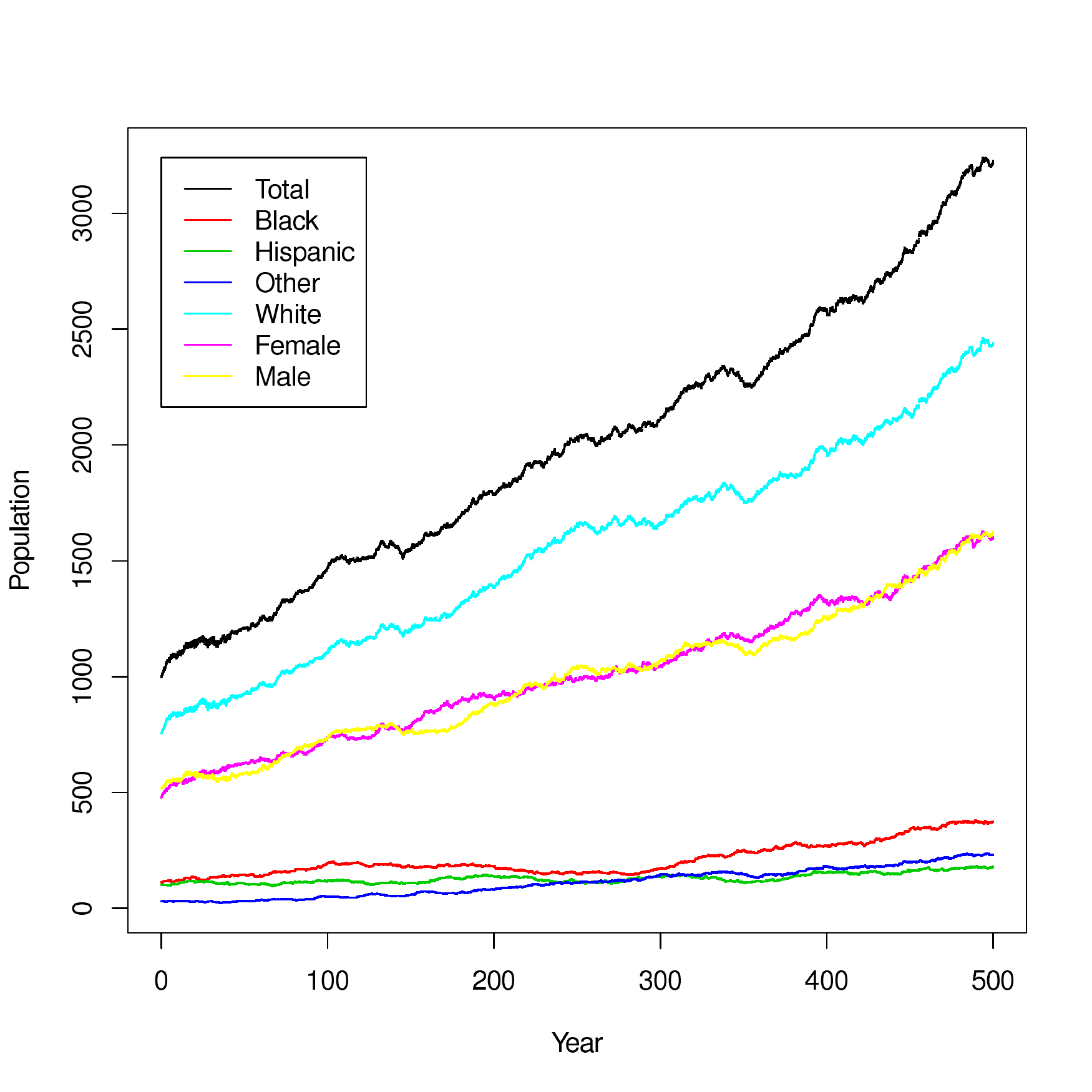}&
          \includegraphics[width=0.48\columnwidth,keepaspectratio]{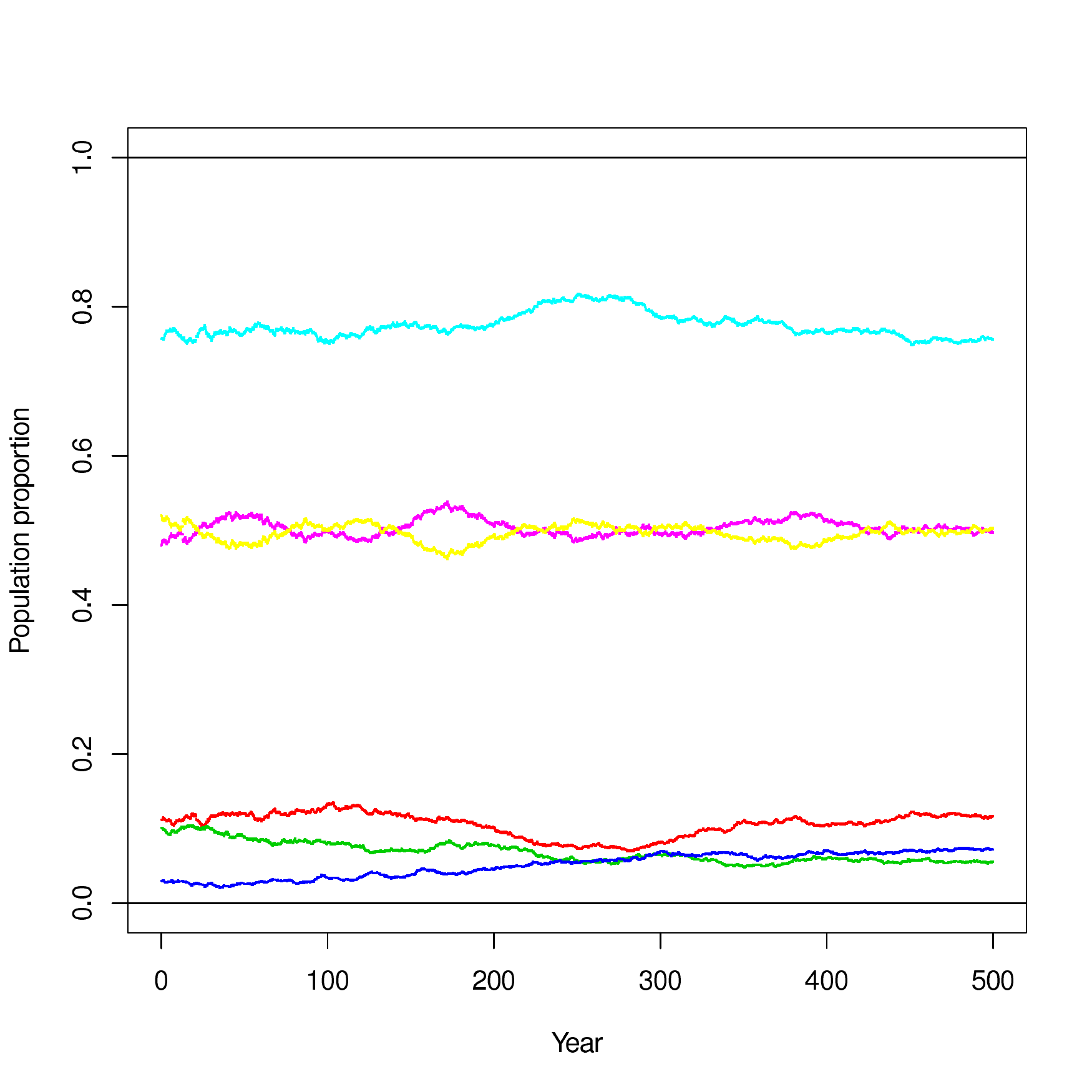}\\
          (a) subpopulation sizes & (b) subpopulation proportions
        \end{tabular}
      \end{center}}
    \caption[Simulated population over time]{\label{fig:NHSLS-ML-n}
      NHSLS dynamic simulation population (network size) and subpopulation sizes and proportions over time.}
  \end{figure}
  
  We give snapshots of
  cross-sectional network statistics at $t=100$, $t=300$, and $t=500$ years in table
  Table~\ref{tab:NHSLS-ML-results}. Plots of trends over time are
  given in Figure~\ref{fig:NHSLS-ML-results}.

\begin{figure} \noindent {\centering
\includegraphics[width=0.95\columnwidth,keepaspectratio]{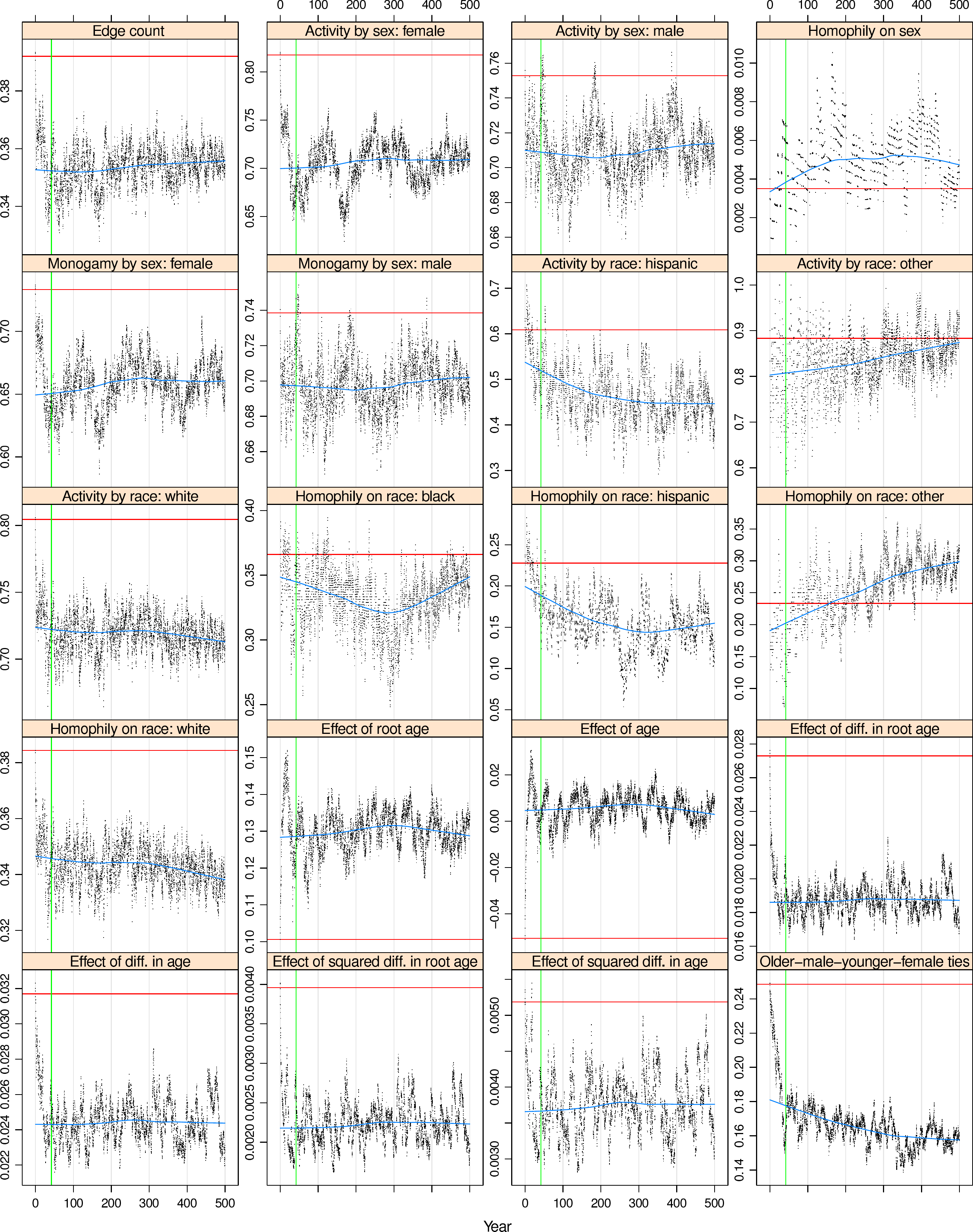}
}
\caption[Simulated network statistics in a growing
population]{\label{fig:NHSLS-ML-results} NHSLS dynamic simulation
  network statistics over time, in a growing population. All values
  are normalized per Section~\ref{sec:NHSLS-model-fit}. Red lines are the \enquote{target} values for their
  respective statistics; green line is the time point at which no
  actors from the original population remain.}
\end{figure}

The simulation of the evolving population removes individuals and
their ties from the network, and adds only individuals with no ties,
something not taken into account by the algorithm used to fit the
parameters, so, after a short \enquote{burn-in} period, the mean
degree of the network is uniformly lower than that targeted. Beyond
that, the mean degree (and other such statistics) do not appear to be
sensitive to network size, but do appear to be sensitive to the
population composition: the trends in racial-category-specific mean
degrees (Figure~\ref{fig:NHSLS-ML-results}, the \enquote{homophily on
  race} and \enquote{activity by race} panels) closely follow those in
the population (Figure~\ref{fig:NHSLS-ML-n}(b)). This is what the
model of \citet{krivitsky2011ans} predicts, assuming the
\enquote{preferences} as modeled remain unchanged: the more rare a
given individual's preferred partners are in the population, the lower
the expected mean degree for that individual.  }
\subsubsection{Duration distribution}
The approach of Section~\ref{sec:implementation} gives a target mean
duration of 10.31 years, but because about 28\% of ties in the history
of the simulated network had been censored due to the actor being
removed from the population of interest and 2\% of ties had been
censored due to the simulation ending, the average duration of a tie
that is simulated is 7.16 years. Using a Kaplan-Meier estimator
\citep{therneau2009ssa} to estimate the survival function $1-F(x)$,
and estimating the mean of duration by integrating it
\citep[117--118]{klein2003sat} gives an adjusted mean duration of
10.18 years, slightly smaller than the target, likely due to no
simulated tie having duration greater than $60-18=42$ years.

All this suggests that this model is at the very least viable for
simulating networks in populations with changing size, based on fairly
limited data.
\section{Discussion}
We began with the model of \citet{KrHa14s}, described its
long-run properties, especially in the simpler and more tractable
special cases; and discussed approaches --- both feasible and not ---
to fitting these network evolution processes to sociometric data that
are not, at first glance, amenable to dynamic network modeling. We
have also described how the network size adjustment of
\citet{krivitsky2011ans} may be incorporated into this model and
demonstrated its efficacy with an application to the NHSLS survey.

Many simplifying assumptions were made in the simulation, aspects left
unanalyzed, and questions unanswered about the properties of the
models discussed. We describe some of them here.

The EGMME approach depends on a very strong assumption, that the
social process being modeled does not significantly change over time
in ways that the model does not account for, so the network data
observed can be plausibly modeled a draw from an equilibrium
distribution. It is not clear whether this assumption can be tested,
especially if only egocentrically sampled data are available, beyond
the generic goodness-of-fit measures like those of
\citet{hunter2008gfs}.  Conditional approaches do not suffer from
this, but, as we showed in Section~\ref{sec:prob-conditioning},
require data not often available, and methods to circumvent this
problem also require fairly strong assumptions, though those
assumptions are arguably weaker than those needed for full equilibrium
inference.

Although we have discussed the effects of dyadic dependence in the
dissolution phase of the process, we have not discussed nor
demonstrated how its parameters might be fit. In particular, in a
sexual partnership network, it is very plausible that monogamous ties
are much more stable than those that are concurrent. This effect can
be modeled in a STERGM with a generative statistic
\[\genstatD_k(\yD,\yat{-1})=\sum_{\iactors} \I{\abs{\yD_i}=1}.\]
But, if there is expected to be monogamy bias in formation as well, if
the only target statistic with information about degree is
\[\target_k(\y)=\sum_{\iactors} \I{\abs{\y_i}=1},\]
the monogamy biases in formation and dissolution may not be
identified. A duration-sensitive target statistic like
\[\target_{k'}(\y,\dotsc)=\frac{1}{n}\sum_{\iactors}\I{\abs{\y_i}=1}\sum_{j\in\y_i}\age(\yij),\]
the average duration of all monogamous ties, used alongside
$\target_k$, may be able to identify the incidence effects from the
duration effects, and can be inferred from short-term relationship
history: all that is needed is to know whether a relationship was
monogamous at the time of the survey and how long the relationship had
lasted.

While the simulation incorporated vital dynamics to demonstrate
invariance to network size, the procedure used to fit the parameters
did not take into account vital dynamics in any way. Taking vital
dynamics into account when learning $\tilde{\curvpar}$ is a subject of
ongoing research. One way to do so may be to simulate the effects of
aging and of actors aging out of population of interest (18--59 in the
example above) or otherwise being removed, then \enquote{resetting}
the age of each removed actor to the age of a neophyte (18 in this
case) breaking all of the actor's ties, and reinserting that actor
into the population. Such a process would have a stationary
distribution, especially if the removal process were at least somewhat
stochastic to render the combined process aperiodic, and it thus could
still be used for simulated EGMME, but its estimates for the expected
values of the target statistic under a particular parameter
configuration would at least partly reflect the vital dynamics.

The GMME approach to network inference can be extended to other
\enquote{inconvenient} data: while the targeted statistics we have
used to date have been network statistics, they do not have to be: for
example, infection tree data may contain information about the
underlying network process: rather than finding that network process
which produces networks having statistics similar to those that have
been observed, it may be possible to find that network process which
produces networks, infection processes on which produce infection
trees with similar features to those observed, as an alternative to
the method of \citet{GrWe11b}.

\section{Acknowlegements}
This work was supported by NIH awards R21 HD063000-01, P30 AI27757,
and 1R01 HD068395-01, NSF award MMS-0851555 and HSD07-021607, ONR
award N00014-08-1-1015, NICHD Grant 7R29HD034957, NIDA Grant
7R01DA012831, the University of Washington Networks Project, and
Portuguese Foundation for Science and Technology Ci\^{e}ncia 2009
Program.  The author would also like to thank David Hunter and Mark
Handcock, as well as the members of the University of Washington
Network Modeling Group, especially Martina Morris and Steven Goodreau,
for their helpful input and comments on the draft.

\bibliographystyle{plainnat}
\addcontentsline{toc}{section}{References}
\bibliography{ERGM-based_models_and_inference_for_dynamic_networks}

\appendix

\section{Finding a Generalized Method of Moments Estimator\label{app:gmme}}

Finding a GMME in our setting presents several unusual challenges. The
objective function \eqref{eq:gmme-obj} can only be estimated by
simulation, necessitating some sort of stochastic
approximation. However, many network processes of interest, such as
sexual partnership networks, evolve slowly, with relationships lasting
months or years, even decades. On the other hand, the plausibility of
the separability assumption improves as the length of each discrete
time step increases \citep{KrHa14s}.  Thus, successive
networks drawn from the model are likely to be very similar --- highly
autocorrelated. This, in turn, means that a sufficiently precise
estimate of $J(\curvpar)$ requires simulating a very long series of
such networks.

Finding the direction of the stochastic search presents a further
challenge. In gradient-based methods,
\[\nabla J(\curvpar) = \left(\meanstats(\curvpar)-\target(\y^1,\dots,\y^T)\right)\t \V(\curvpar)^{-1} \gradient(\curvpar). \]
Some specific combinations of $\target$ and $\genstat$ suggest some
simple relationship for $\gradient$. For example, if
$\target_k(\yat{})\equiv\abs{\yat{}}$ and
$\genstat_l(\yat{},\yat{-1})\equiv\abs{\yat{}\cup\yat{-1}}$ (i.e.,
edge count formation), it is likely that the
$\gradient_{k,l}(\curvpar)>0$. However, for less strongly related
statistics, the gradient would need to be estimated.

Nor are the signs of elements of the $\gradient(\curvpar)$ guaranteed
to remain the same throughout the parameter space. For example, if
$\target_k(\yat{})\equiv\sum_{i = 1}^\nactors \I{\abs{\yat{}_i=1}}$
--- the number of actors with degree 1 --- and $\genstat_l$ is as
before, then the gradient is likely to be positive when $\curvpar_l$
is low, and most actors are isolates, since an increase in
$\curvpar_l$ increases the number who have one tie; but as
$\curvpar_l$ increases to the point where some actors begin to acquire
their second tie, its effect on the number of actors with one tie
reverses, to the point where increasing $\curvpar_l$ makes it
increasingly less likely that an actor will have fewer than two
ties. Thus, the gradient matrix must be estimated and reestimated
continually throughout the search.

Selection of starting values for the optimization presents yet another
challenge. As with ordinary ERGMs, a poor choice of starting parameter
configuration may induce extreme network distributions, in the sense
that $\meanstats(\curvpar_0)$ is close to the edge of the convex hull
of possible network statistics, which, in the discrete space of
networks, makes it almost impossible to estimate
$\meanstats_k(\curvpar_0)$, because almost all equilibrium draws
$\target_k(\Yat{})$ under $\curvpar_0$ equal to
$\min_{\ynetsY}\target_k(\y)$ or $\max_{\ynetsY}\target_k(\y)$.
Unlike ordinary ERGMs, where the Maximum Pseudolikelihood Estimator
(MPLE) provides a plausible $\curvpar_0$, we are aware of no such
methods. Thus, the optimization method must be relatively robust to
poor starting values. In particular, if the initial configuration has
some expectations be near their minimal or maximal values but not
others, the algorithm should not necessarily fail: perhaps finding a
parameter configuration where $\meanstats_k(\curvpar)$ is close to
$\target_k(\y^1,\dots,\y^T)$ for some $k$ will shift others to a more
convenient region. An example of this is
$\target(\yat{})\equiv(\abs{\yat{}},\sum_{i = 1}^\nactors
\I{\abs{\yat{}_i=1}})\t$: that if that an initial parameter
configuration inducing overly dense networks can cause
$\meanstats_2(\curvpar_0)\approx 0$, but if further optimization using
$\meanstats_1$ brings the network density down to where more actors
have one tie, the estimation could begin to incorporate the second
statistic.

We have tried a number of approaches, including Kiefer-Wolfowitz
\citep{kiefer1952sem} and Simultaneous Perturbation Stochastic
Approximation (SPSA) \citep{spall1998osp}, but we have ultimately
found that simple gradient descent updating, with
\[\curvpar^{t}=\curvpar^{t-1}-\gamma_t \left(\hat{\meanstats}(\curvpar^{t-1})-\target(\y^1,\dots,\y^T)\right)\t \hat{\V}(\curvpar^{t-1})^{-1} \hat{\gradient}(\curvpar^{t-1})\]
works adequately, with $\gamma_t$ being a declining sequence, $\meanstats(\curvpar^t)$ estimated by simulation,
$\gradient(\curvpar^t)$ being estimated by regressing recent values of
$\target(\Yat{})$ on $\curvpar^t$ (and an intercept). The covariance
of the residuals from this regression is then used to estimate
$\V(\curvpar^t)$. The resulting process produces a \emph{Continuous-Updating Estimator}. \citep{hansen1996fsp}

\end{document}